# Mathematical Characterization of Thermo-reversible Phase Transitions of Agarose Gels

Arif Selcuk Ogrenci [a], Onder Pekcan [a], Selim Kara [b], Ayse Humeyra Bilge [a]
ogrenci@khas.edu.tr, pekcan@khas.edu.tr, skara@trakya.edu.tr, ayse.bilge@khas.edu.tr
[a] Faculty of Engineering and Natural Sciences, Kadir Has University, 34083, Istanbul, Turkey
[b] Department of Physics, Trakya University, 22030, Edirne, Turkey

**ABSTRACT**
Thermal phase transition temperatures of high and low melting point agarose gels were investigated by using the UV–vis spectroscopy techniques. Transmitted light intensities from the gel samples with different agarose concentrations were monitored during the heating (gel-sol) and cooling (sol–gel) processes. It was observed that the transition temperatures $T_m$, defined as the location of the maximum of the first derivative of the sigmoidal transition paths obtained from UV–vis technique, slightly increased with the agarose type and concentration. Here, we express the phase transitions of agar-water system, as a representative of reversible physical gels, in terms of a modified Susceptible-Infected-Susceptible epidemic model whose solutions are the well-known 5-point sigmoidal curves. The gel point is hard to determine experimentally and various computational techniques are used for its characterization. Based on previous work, we locate the gel point $T_0$ of sol-gel and gel-sol transitions in terms of the horizontal shift in the sigmoidal transition curve. For the gel-sol transition (heating), $T_0$ is greater than $T_m$, i.e. later in time, and the difference between $T_0$ and $T_m$ is reduced as agarose content increases. For the sol-gel transition (cooling), $T_0$ is again greater than $T_m$, but it is earlier in time for all agarose contents and moves forward in time and gets closer to $T_m$ as the agarose content increases.

1. **INTRODUCTION**

In general, gels are classified by the type of the cross-linkages. Some gels are cross-linked chemically by covalent bonds, while others are cross-linked physically either by hydrogen or ionic bonds [1, 2]. Since they cannot be dissolved again, gels formed by chemical bonds are irreversible gels. On the other hand, physical gels are reversible under heating and cooling processes. There are various ways in which the monomers can be arranged in the polymer chain. In condensation gelation, the molecules cross-link into larger clusters by forming covalent bonds in several ways [3]. Free-radical cross-linking copolymerization (FCC) has been widely used to synthesize polymer gels. A chemical gel appears during a random cross-linking process of monomers to larger and larger molecules. On the other hand, moderate heating can reversibly dissolve a physically cross-linked gel, which can be named as weak gels. Then, the process of gelation upon cooling is a sol–gel phase transition and the reversible gelation process is called a gel-sol phase transition [4]. Gel's phase transitions can be affected by solvent composition, pH changes, ion composition changes, and applied electric field. Many of the natural polymer gels fall into the class of physical gels, among which red algae attracted attention for various applications. Red algae produce a wide range of galactose based polysaccharides, one of which agar has attracted great interest because of its applications in food and other industries.

Agar and its derivative agarose are extracted from certain red algaes, and agar consists of agarose and agaropectin fractions [5]. Agarose is a neutral and linear polysaccharide, and forms thermoreversible gels when dissolved in water. Agaropectin is a mixture of charged and sulfated non-gelling galactans.

The common feature of certain types of biopolymers such as agarose, carrageenan and gelatine is their ability to form a gel structure in water. Since those hydrogels possess the properties of volume and thermal phase transitions, as well as a large amount of solvent absorption, they have remained attractive for experimental and theoretical researchers. Agarose is one of the most popular polysaccharides used in molecular biology and bio-technological applications [6, 7]. It is also widely utilized in medicine, cosmetics and food industries because it forms strong gels even with lower agarose concentrations [8, 9]. Gel formation in an aqueous solution of gel forming polysaccharides is a complex process that depends on the polysaccharide structure, polymer concentration, temperature and some specific counterions.

The gelation and thermoreversible sol–gel processes of biomacromolecules involve intra- and inter-molecular hydrogen bondings and electrostatic and hydrophobic interactions leading to different supramolecular structures, and therefore are of high intrinsic interest [10, 11]. Unlike anionic carrageenans, agarose is essentially a neutral polymer that requires no counterions or other additives to induce gelation [12].

The gelation processes of agarose and agarose-like biogel systems (e.g. carrageenans) have been widely studied via experimental and theoretical techniques over the last several decades to produce specific properties for specific applications. For example, the kinetics and equilibrium processes of the sol-gel transitions of agarose gels as well as the effect of gelation conditions on the gel's microstructure and rheological properties like the effect of salts and ions [8,13], influence of thermal history [14] and pore-size determination [9, 15] have been studied in recent years.

A widely used model for the gelation mechanism of agarose was proposed by Tako and Nakamura who built the model upon previous experimental results [16]. When the agarose/water system which is in the sol state at higher temperatures (60 °C) started to cool, it was found that intramolecular hydrogen bonds are established and form a single helix structure. In consequence of this event, the single helical agarose chains become more rigid, and a further decrease in temperature results in intermolecular bonds between those single helices, leading to double helical formations [17]. At lower temperatures, assembling of the double helices creates double helix aggregates in the gel network. Water molecules are also bonded to the agarose chains, contributing to the total rigidity of the chain.

However, other studies related with agarose network formation propose that the agarose network consists of junctions of substantial bundles of agarose chains. According to this network model, crosslinks involve both the double helix formation and substantial association of the double helices to form microcrystalline junction zones [17].

The aim of this study is to investigate the thermoreversible phase transitions of two kinds of agarose gel systems at the molecular level by using photon transmission techniques. Transmitted light, $I_{tr}$, intensities were measured against temperature to monitor the sol-gel and gel-sol phase transitions in order to determine the transition temperatures $T_m$ and $T_0$. Here, $T_m$ was produced from the position of the first derivative of $I_{tr}$ with respect to temperature. $T_0$ is defined as the gel point for sol-gel and gel-sol phase transitions.

## 2. THE GEL POINT

In general, the gel point can be identified using two different techniques. One of which, for the gel point measurements, dilatometers containing a steel sphere were used, where the midpoint between the last time at which the sphere moves magnetically and that at which it stops moving is taken as the gel point. In dilatometric technique [18] for determining the gel points, each experiment was repeated at the same experimental conditions, and the gel points were determined by a steel sphere which was moved in the sample up and down slowly by means of a magnet applied to the outer face of the sample cell. The time at which the motion of the sphere was stopped was evaluated as the gel point. The steel sphere in the samples of higher polymer concentration cannot be moved after the polymerization is complete, but it can partly be vibrated around its equilibrium position for the loosely formed gels, by moving the magnet up and down on the surface of the sample cell. We reported previously the consistency in the gel points determined using dilatometric and gravimetric methods [19].

On the other hand, in the language of percolation, one may think of monomers as occupying the sites of a periodic lattice, and of the chemical bonds as corresponding to the edges joining these sites randomly with some probability p. Then the gel point can be identified with the percolation threshold $p_c$, where, in the thermodynamic limit, the incipient infinite cluster starts to appear. During gelation one would like to measure the values of the critical exponents γ and β with sufficient accuracy to determine their universality class and to verify that they indeed have the non-classical values for percolation, computed from series expansions and Monte Carlo studies as well as renormalization theory. Here the exponents γ and β are for the weight-average degree of polymerization, $DP_w$, and the gel fraction, G as given below.

$$DP_w = A(p_c - p)^{-\gamma} \qquad (1)$$

$$G = B(p - p_c)^{\beta} \qquad (2)$$

The double logarithmic plot of the measured quantity against $|p - p_c|$ gives a critical exponent as the slope of the straight line. Fitting the data, a main obstacle lies in the precise determination of the critical point and/or gel point at the critical region. In particular, a small shift in $p_c$ results in a large shift in the critical exponent. Such a log–log plot reveals that data should be particularly accurate near the gel point. Preferably, one should have more than one quantity measured in the gelation experiments. Then, one can fix $p_c$ from the best fit of data and use the same $p_c$ for other properties [20]. The way to find the critical point in real experiments is to measure and to perform the scaling analysis for more than one quantity [21]. The critical point can then be determined by varying $p_c$ in such a way as to obtain good scaling behaviour for both quantities over the greatest range in $|p - p_c|$, or $|t - t_c|$ if the experiments are performed against time. The first time in 1998, it has been reached the experimental state of art to determine exponents for the poly(methyl methacrylate) (PMMA) by means of the steady state fluorescence (SSF) technique.

The determination of the gel point in phase transitions has been an intriguing problem because its location with respect to the inflection point of the sigmoidal transition curve depends on the nature of the gelation experiments. The gel point, $T_0$ is characterized by a drastic change in various physical properties and it is hard to measure without disturbing the transition process. As an effort to characterise the gel point based on nonintrusive, spectroscopy based light intensity measurements, the gelation process has been modelled in terms of the epidemic dynamics where systems of differential equations governing SIR (Susceptible-Infected-Removed) and SEIR (Susceptible-Exposed-Infected-Removed) models have been utilized [22]. In experimental work that provides the basis of the results presented in [22], the gel points were determined by independent experiments. After successfully modelling these gelation processes in terms of the SIR and SEIR systems of differential equations, we were challenged by finding a mathematical property that corresponds to the gel point. We observed that the derivatives of the sigmoidal curve that represent the Removed individuals of the SIR and SEIR models have an interesting property. The time instants $t_i$ at which the i'th derivative reaches its global extremum formed a seemingly convergent sequence. This limit point turned out to be in qualitative agreement with the location of the gel point as measured in [22] and the existence of such a limit point was proposed as a mathematical definition of the critical point of a phase transition [23]. Later on we proved that, the existence of a critical point of a sigmoidal curve, in the sense above, was due to the wave packet behaviour of the derivatives and it could be expressed in terms of the properties of the Fourier transform of its first derivative [24]. We have also computed the Fourier transform of the generalized logistic growth [25] and located its critical point.

In another work, we had observed that the SIR and SEIR models were inadequate for modelling the gelation of K-carrageenan-water systems [26], which are typical examples of reversible, physical gelation processes. In the present work we model these reversible physical gelation processes by a modification of the Susceptible-Infected-Susceptible (SIS) epidemic model, as discussed in Section 4. It is well known that the SIS model is equivalent to the generalized logistic equation and its solutions can be expressed in terms of the standard logistic growth curve [22]. Thus, data and the solution curves of the model can be compared directly by using regression software. However, the gelation data of [26] does not fit the standard logistic growth; but it fits the generalized logistic growth almost perfectly. Accordingly, in Section 4, we introduce a new parameter to obtain a modified SIS system whose solutions are generalized logistic growth curves. This result allows interpreting the chemical and physical gels in a common framework, in terms of epidemic models.

In Section 5, we use these results to locate the gel point in the sol-gel and gel-sol transitions, as the critical point of the generalized logistic growth curve that best fits the data.

### 3. EXPERIMENTAL WORK

Eight samples of HMP (High Melting Point) and seven samples of LMP (Low Melting Point) agarose gels at different concentrations were prepared. Agarose powder was slowly added to 10 ml of distilled water while stirring and heating the mixture up to 95 °C for 20 min. Amount of HMP powder was between 30 mg and 400 mg, and LMP powder was between 50 mg and 300 mg. The UV-visible measurements were performed using Varian Cary Eclipse spectrophotometers equipped with temperature controller units. The UV–vis data was collected at a wavelength of 400 nm in photon transmission intensity measurements. The transmitted photon intensities, $I_{tr}$, were monitored against temperature in the range of 15 to 95 and 15 to 80°C for the HMP and LMP samples, respectively.

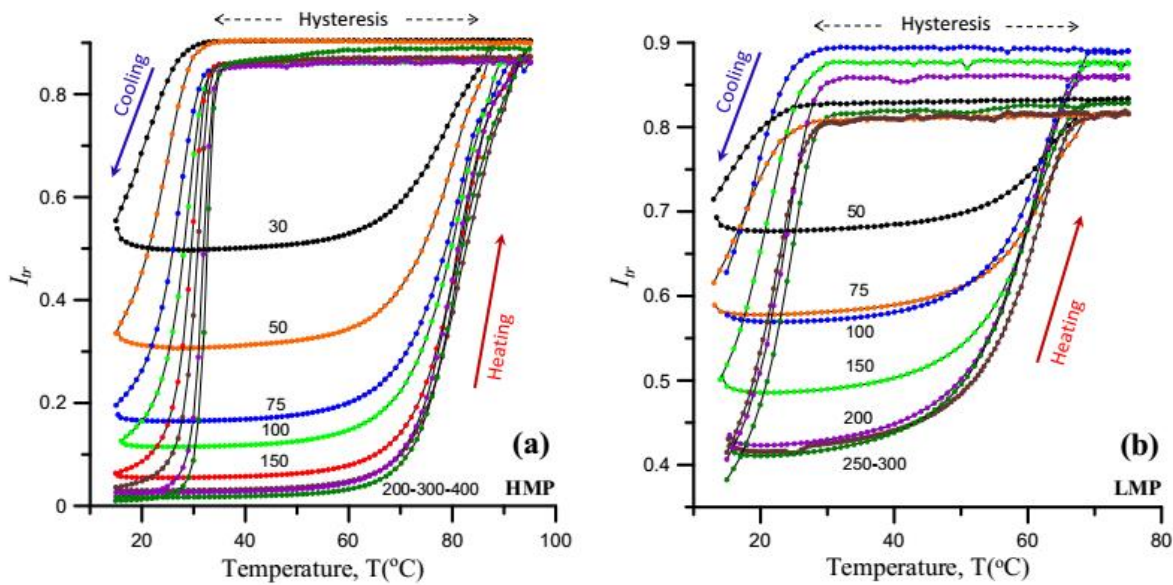

Figure 1. Transmission intensity measurements during sol-gel (cooling) and gel-sol (heating) transitions with different agarose content for the a) HMP and b) LMP samples.

Figure 1 shows the temperature variations of $I_{tr}$ intensities for the HMP and LMP agarose samples. It can be seen from the HMP curves in Fig. 1a, that the $I_{tr}$ shows a slight increase in the low temperature region (15–65 °C) and a more radical increase in the high temperature region (65–95 °C) during heating. The LMP samples in Fig. 1b, indicate the same behaviour for the lower (15–50 °C) and the higher (50–80 °C) temperature regions. At the beginning of the heating process (lower temperature region), the gel structure consists of double helix aggregates and water is trapped in the gel network. Thus, the incident light scatters in different directions through the heterogeneous gel structure with different indices of diffraction. This prevents the penetration of photons into the gel, resulting in lower $I_{tr}$ values. In the high temperature region, however, the amplitudes of molecular vibrations start to increase, resulting in the destruction of the double helix aggregates, and subsequently the gel-sol phase transition is accomplished by the decomposition of the weak hydrogen bonds which constitute the double helices, and the agarose chains are homogeneously dispersed in the water. As the agarose-water system becomes homogeneous, the scattered light intensity decreases, and the system becomes fully transparent, i.e. the $I_{tr}$ rapidly increases (see Fig. 1a). In the cooling process, the system starts to reassemble the hydrogen bonds which are easily disrupted by the effect of the increasing temperature. This is possible only when the decreasing amplitudes of the molecular vibrations allow the agarose molecules to take the proper positions to bond to each other. Since this can happen at lower temperatures, the $I_{tr}$ show no distinct variation until 30 °C (see Fig. 1a). As a consequence of this delay, hysteresis occurs between the heating and cooling processes. This behaviour is a common characteristic of polysaccharides which can perform thermoreversible phase transitions. The main cause of hysteresis is the difference between the processes of the formation and the destruction of the gel network which consists of double helices and their aggregates [27, 28]. As the temperature decreases below 30 °C, the agarose chains take the shape of helix formations (intramolecular hydrogen bonding). Then, the double helix structures (intermolecular hydrogen bonding) and aggregates are rapidly constituted by appropriate positioning of the double helices. In other words, double helices are formed through the association of agarose molecules and then the double helices are aggregated to higher ordered assemblies to create a three dimensional network. During gelation, the agarose-water system starts to form two-phases with different network concentrations, which creates concentration fluctuations. Namely, the double helix aggregates are formed as a separate phase by excluding water from their domains. As a result, the contrast between the agarose and water phases plays an important role in scattering the light. As these explanations indicate, the gel structure becomes more heterogeneous for the light and $I_{tr}$ falls to minimum levels. The increase in the strength of scattering causes a decrease in the number of the photons which penetrate into the sample and excite the pyranine molecules. Arguments similar to given above can be repeated for the LMP samples, as shown in Fig. 1b, during the heating and cooling processes for $I_{tr}$ intensities. In Fig. 1, it can be seen from the $I_{tr}$ variations of LMP and HMP samples with the same concentrations that the gel structures of the samples completely disintegrate around 65 and 90 °C, respectively. So, it can be surmised that the density of the helices and aggregates in the LMP agarose gels is lower than that of the HMP samples.

The transition temperatures, $T_m$, can be produced from the inflection points of the sigmoidal curves of thermal phase transitions [22, 29]. The sol-gel and gel-sol transition temperatures ($T_m$) were determined from the peak positions of the first derivatives of the $I_{tr}$ curves in Fig. 1. The variations of the $T_m$ values compared to different HMP and LMP agarose concentrations are shown in Fig. 2.

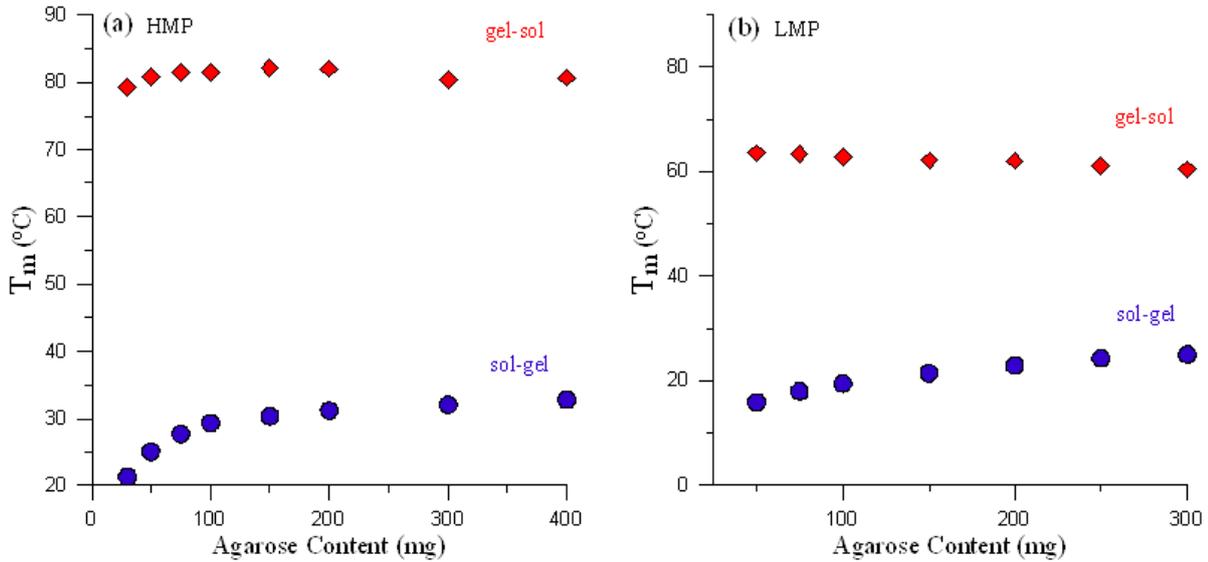

Figure 2. Variation of the critical temperature Tm in sol-gel (cooling) and gel-sol (heating) transitions with respect to the agarose content for the a) HMP and b) LMP samples.

As can be seen from Fig. 2, the increase in the agarose content up to 150 mg causes a significant increase in the sol-gel transition temperatures. Further increasing the concentration leads to saturation of the sol-gel transition temperatures. It should also be noted that the gel-sol transition temperatures do not show a distinct increase in comparison to the agarose concentrations. It is clear that the gel-sol critical transition temperatures are much higher than the sol-gel temperatures for the both of the agarose types, which is in agreement with our previous findings on k-carrageenan and agarose gel systems [30, 31]. The difference between the transition temperatures is a result of the hysteresis which occurs during the thermoreversible gelation process. In other words, the formation of the gel structure is possible at lower temperatures; however, the dissociation of the gel structure requires higher temperatures. Our findings are in agreement with the work of Arnott et al. [32]. On the other hand, in our study the measured $T_m$ values of the low molecular weight (LMP) gel samples are 10 and 20 °C lower than that of the high molecular weight (HMP) gel samples, for the sol-gel and gel-sol transitions respectively. As it has been noted before, the LMP agarose produces weaker gels through the effect of the lower density of helices and their aggregates. HMP samples have denser structures than that of LMP samples [33]. For this reason, the association and dissociation temperatures of the LMP gel requires lower values with respect to the HMP samples.

## 4. THE MATHEMATICAL MODEL

In a series of papers we studied the gelation process in terms of epidemic models. We have shown that the formation of chemical gels can be represented by the Susceptible-Infected-Removed (SIR) epidemic model [22]. During this study, we observed that the absolute extrema of the higher derivatives agglomerate at a certain point and motivated by this fact we defined the critical point of a sigmoidal curve as the location of the limit point of the points where the derivatives reach their absolute extreme values [24], as shown in Figure 3. We have shown that, for chemical gels, the location of the mathematically determined critical point agrees qualitatively with the location of experimentally measured gel point [25]. Recently, we modelled the sol-gel and gel-sol transitions of physical gels in terms of a modification of the Susceptible-Infected-Susceptible (SIS) epidemic model [34]. The solution of the modified SIS model turns out to be the generalized logistic growth curve for which the location of the critical point is the parameter $x_0$ given below [24, 25].

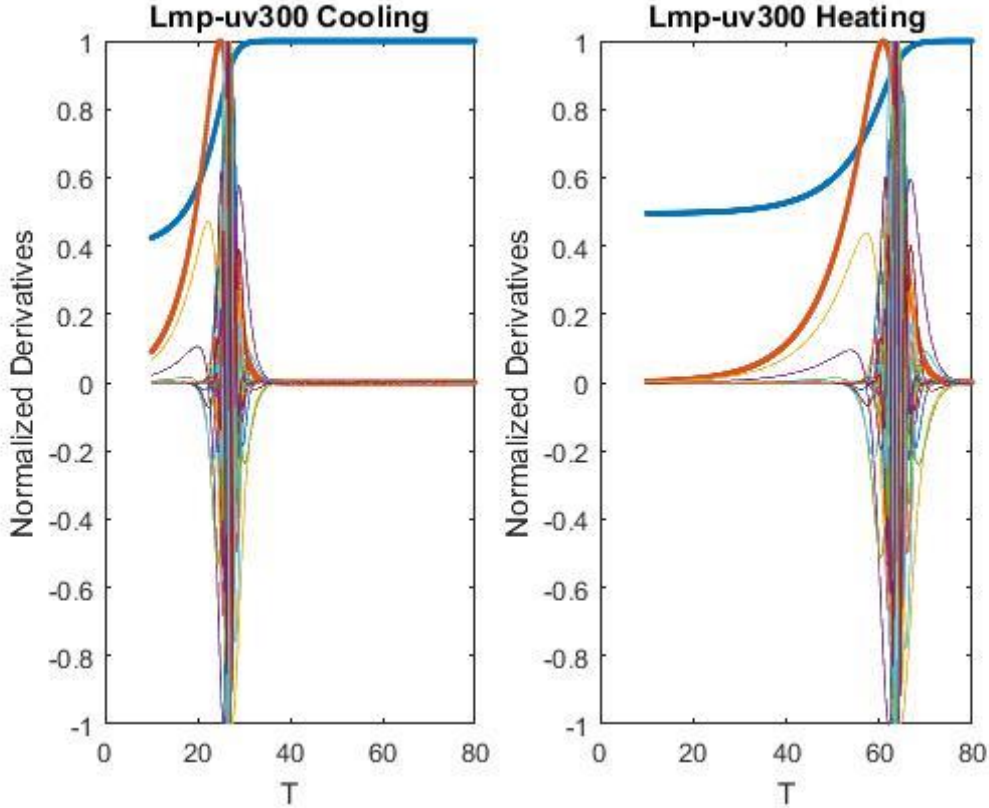

Figure 3. Transition curve and its derivatives for two samples where the absolute extreme values of derivatives converge to the critical temperature, a) LMP-uv300 cooling and b) LMP-uv300 heating.

The generalized logistic growth is governed by the equation

$$y = y_0 + a\,[1 + e^{(x-x_0)/b}]^{-c} \qquad (3)$$

In this equation the parameters b and c are crucial in determining the shape of the sigmoid. The location of the maximum of the first derivative is obtained by finding the (unique) zero of the second derivative. It is given by

$$x_m = x_0 + b\,\ln(c) \qquad (4)$$

The locations of the inflection points of the first derivative are also important in determining the shape of the sigmoidal curve. These are obtained as the zeros of the third derivative, computed as

$$x_1 = x_0 + b\,\ln(u_1), \qquad x_2 = x_0 + b\,\ln(u_2) \qquad (5)$$

where

$$u_1 = (1/2)\,[(3c+1) - (5c^2+6c+1)^{1/2}], \qquad u_2 = (1/2)\,[(3c+1) + (5c^2+6c+1)^{1/2}] \qquad (6)$$

Here, clearly, $x_1 < x_2$.

We give below the graphs of the sigmoid and its first derivative for typical values of c and for b=2. For Figure 4, below, with c=1, the first derivative is an even function and the points $x_m$ and $x_0$ coincide. For c>1, the critical point $x_0$ is located before the maximum of the first derivative $x_m$. For c<1, the critical point $x_0$ is located after the maximum of the first derivative as can be seen in Fig. 4.

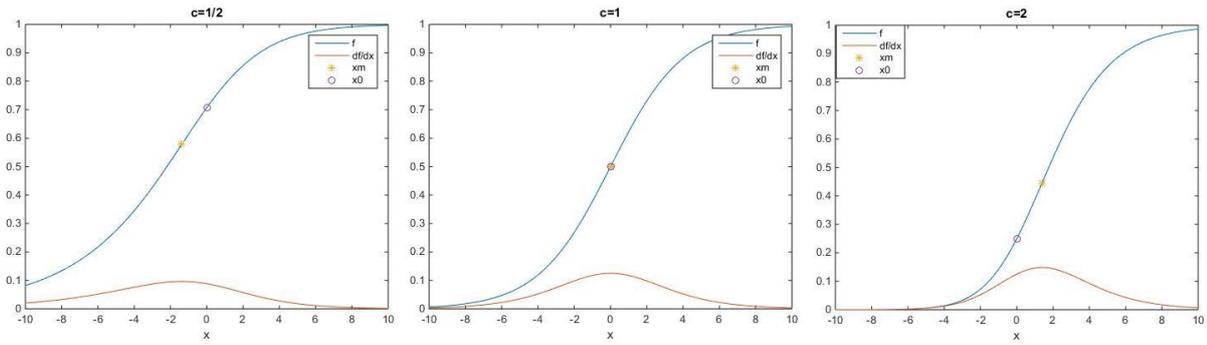

Figure 4. Location of the transition point $x_m$ and the critical point $x_0$ in the sigmoid curve for different values of c=0.5, 1, 2.

The relative location of the critical point and the inflection point is an invariant measure of the closeness of the critical point $x_0$ to the transition point $x_m$ and to the inflection points $x_1$ and $x_2$. It can be seen that, their relative locations are as given below for the possible values of the parameter c.

$$
\begin{aligned}
&0<c<1 & &x_1 < x_m < x_0 < x_2 \\
&c=1 & &x_1 < x_m = x_0 < x_2 \\
&1<c<3 & &x_1 < x_0 < x_m < x_2 \\
&3<c & &x_0 < x_1 < x_m < x_2
\end{aligned}
\quad (7)
$$

These points are shown in Fig. 5. We also define the following ratios that can be used as a measure of the dispersion of the critical point with respect to the inflection points.

$$K^+ = (x_0-x_m)/(x_2-x_m) = -\ln(c)/(\ln(u_2/c)$$
$$K^- = (x_m-x_0)/(x_m-x_1)) = -\ln(c)/(\ln(u_1/c) \quad (8)$$

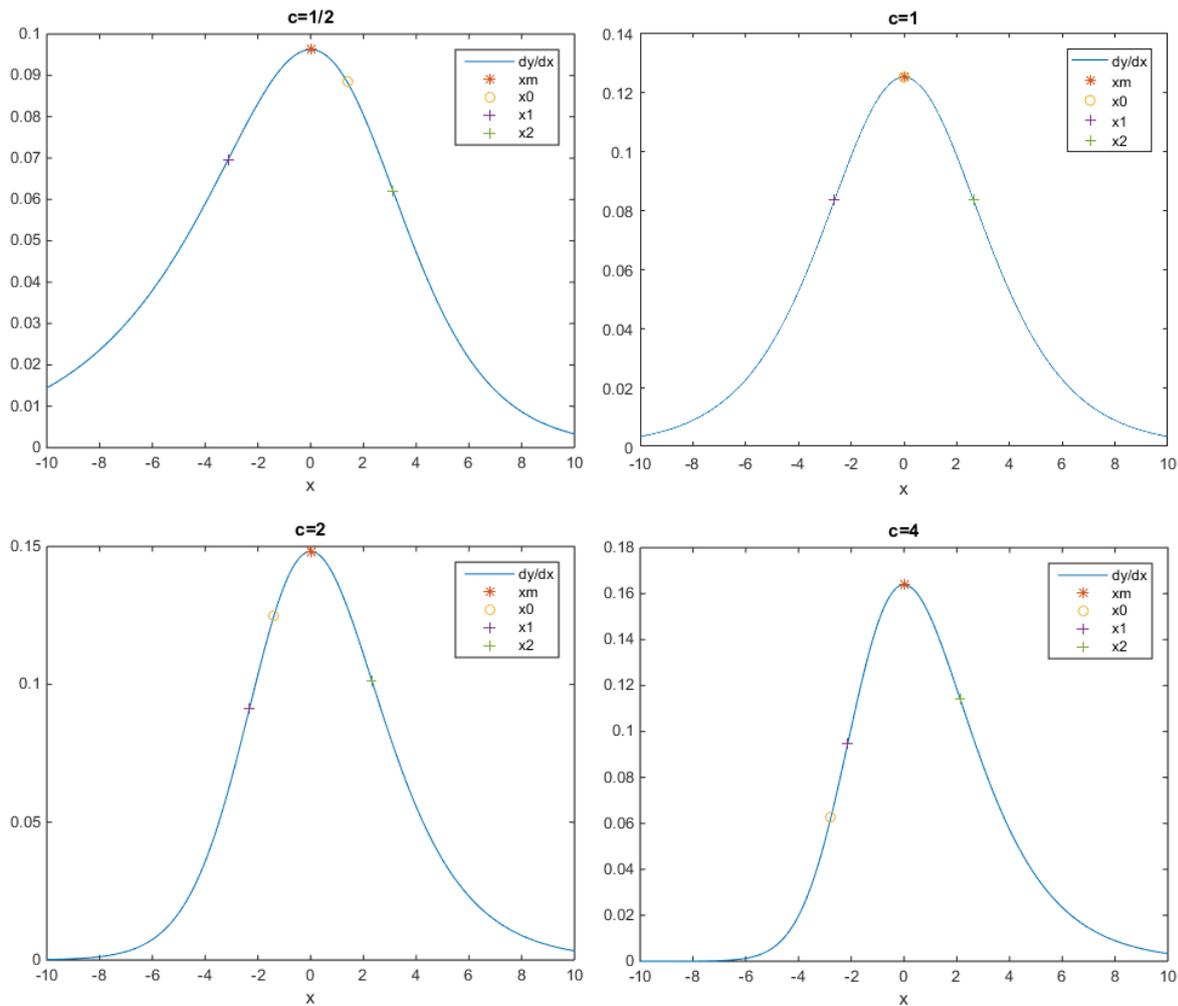

Figure 5. Relative location of $x_0$ and $x_m$ with respect to the inflection points for different values of c=0.5, 1, 2, 4. (The arrows indicate the distances of the critical points and the inflection points to the transition point.)

For 0<c<1, the critical point is located in between the second inflection point and the maximum. For 1<c<3, it is between the first inflection point and the maximum. But if c>3, the critical point is before the first inflection point.

## 5. RESULTS and CONCLUSION

Based on the mathematical model developed above, the measured sol-gel and gel-sol transition data for the two types of agarose, namely LMP and HMP, have been fitted to the generalized logistics growth curve for each of the agarose content separately. A nonlinear regression software has been utilized for this purpose. The fitting performance can be validated by the $R^2$ term (variance explained, or correlation between the measured values and the fitted values) which takes values above 0.998 indicating a good value of fitting performance. The parameters of the sigmoid curve obtained are given in Table 1. for each of the measurement data set, having subclasses with respect to the type of the gel (LMP or HMP), agarose content, and type of process (cooling or heating).

The value of the parameter c remains less than 1 for all the data sets except the two instances at the boundaries. Hence, the expected critical point $T_0$ will be between the transition point $T_m$ and the second inflection point $T_2$ as described above. The location of $T_0$ and $T_m$ are plotted below for LMP and HMP agarose gels in Fig. 6.

**Table 1.** Parameters of the sigmoid curve obtained using the curve fit software.

| Data from | a | b | c | $x_0$ | $y_0$ |
|---|---|---|---|---|---|
| Lmp-uv050 cooling | 0.1378 | 3.5564 | 8.7335 | 7.8318 | 0.6932 |
| Lmp-uv075 cooling | 0.2524 | 2.8753 | 0.7498 | 18.2892 | 0.5596 |
| Lmp-uv100 cooling | 0.4259 | 1.8898 | 0.2515 | 22.1965 | 0.4671 |
| Lmp-uv150 cooling | 0.4847 | 1.7612 | 0.2623 | 24.0212 | 0.3919 |
| Lmp-uv200 cooling | 0.4760 | 1.7024 | 0.3166 | 24.6486 | 0.3368 |
| Lmp-uv250 cooling | 0.5367 | 1.3214 | 0.2151 | 26.3720 | 0.3225 |
| Lmp-uv300 cooling | 0.4978 | 1.3196 | 0.2378 | 26.7594 | 0.3239 |
| Lmp-uv050 heating | 0.1562 | 1.4800 | 0.1811 | 67.2775 | 0.6785 |
| Lmp-uv075 heating | 0.2387 | 1.5821 | 0.1950 | 66.5160 | 0.5791 |
| Lmp-uv100 heating | 0.3203 | 1.6362 | 0.2214 | 65.7997 | 0.5702 |
| Lmp-uv150 heating | 0.3895 | 1.5746 | 0.2054 | 65.2455 | 0.4873 |
| Lmp-uv200 heating | 0.3974 | 1.2784 | 0.1511 | 64.9865 | 0.4174 |
| Lmp-uv250 heating | 0.4373 | 1.5831 | 0.1906 | 64.2109 | 0.4232 |
| Lmp-uv300 heating | 0.4201 | 1.8349 | 0.2129 | 63.7006 | 0.4101 |
| | | | | | |
| hmp-uv030 cooling | 0.6313 | 2.2034 | 0.1738 | 25.2047 | 0.2727 |
| hmp-uv050 cooling | 0.6718 | 1.6388 | 0.2392 | 27.5327 | 0.2302 |
| hmp-uv075 cooling | 0.6958 | 1.3854 | 0.2931 | 29.3374 | 0.1652 |
| hmp-uv100 cooling | 0.7464 | 1.1099 | 0.2887 | 30.5678 | 0.1142 |
| hmp-uv150 cooling | 0.7997 | 0.8722 | 0.3064 | 31.2363 | 0.0671 |
| hmp-uv200 cooling | 0.8258 | 0.7156 | 0.3097 | 32.0754 | 0.0415 |
| hmp-uv300 cooling | 0.8355 | 0.6848 | 0.4299 | 32.6514 | 0.0236 |
| hmp-uv400 cooling | 0.8663 | 0.6606 | 0.5155 | 33.1311 | 0.0143 |
| hmp-uv030 heating | 0.4122 | 3.1117 | 0.3621 | 83.0897 | 0.5011 |
| hmp-uv050 heating | 0.5995 | 2.5930 | 0.3111 | 83.8649 | 0.3103 |
| hmp-uv075 heating | 0.7059 | 2.9387 | 0.3765 | 84.1062 | 0.1664 |
| hmp-uv100 heating | 0.7591 | 2.3889 | 0.3101 | 84.8490 | 0.1156 |
| hmp-uv150 heating | 0.8297 | 3.0836 | 0.4820 | 84.4516 | 0.0566 |
| hmp-uv200 heating | 0.9110 | 4.6031 | 0.8713 | 83.4639 | 0.0319 |
| hmp-uv300 heating | 0.8436 | 3.2579 | 0.6199 | 82.6326 | 0.0277 |
| hmp-uv400 heating | 0.9019 | 4.5482 | 1.0808 | 81.0765 | 0.0191 |

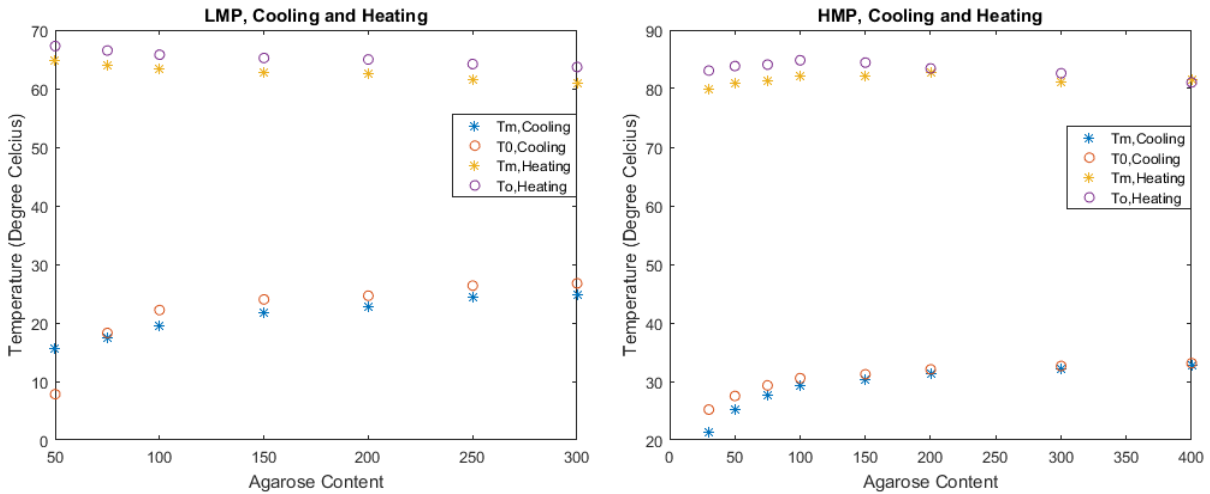

Figure 6. Location of the critical gel point $T_0$ and the transition point $T_m$ with respect to the agarose content for LMP and HMP in sol-gel (cooling) and gel-sol (heating) transitions.

As expected from the mathematical relationship, the critical gel points are slightly above (at a higher temperature than) the transition points except the two boundary cases.

Based on the UV-vis spectroscopy measurements for the sol-gel and gel-sol transition process of HMP and LMP agarose gels, a mathematical model is proposed for the determination of the gel point. The model is based on the numerical observations that the process can be represented using the SIS epidemic model. This leads to the fact that generalized logistics growth curve can be used to obtain a perfect fit for the measurement data. Previous results allow us to identify the critical gel point during the transition process based on the coefficients of the fitted curve provided that the curve is a five point sigmoid which is the case here. Therefore, the data have been fitted using nonlinear regression software, and the critical points have been determined based on the coefficients. Furthermore, the relative positions of the critical temperature $T_0$, transition point $T_m$, and the inflection points $T_1$ and $T_2$ have also been derived for different possible outcomes of the coefficients. For the gel-sol transition (heating), the critical temperature $T_0$ is calculated to be greater than $T_m$, but relatively closer to $T_m$. It can be said that the critical point is later in time for low agarose contents and moves backward as agarose content increases. A similar observation is valid for the sol-gel transition (cooling), where $T_0$ is again greater than $T_m$, and gets closer to $T_m$ as the agarose content increases.